# Néel Skyrmion interactions derived by their arrangement in regular lattices

Ioanna Karagianni, Ioannis Panagiotopoulos*

*Department of Materials Science and Engineering, University of Ioannina 45110, Greece*

*Corresponding author. E-mail address: ipanagio@uoi.gr

**Abstract**

The coarse-grained interactions between Néel skyrmions, stabilized by interfacial Dzyaloshinskii-Moriya coupling (iDMI), are studied through the properties of their hexagonal lattices by micromagnetic simulations. The interactions with the film edges are excluded by imposing periodic boundary conditions. The dependence of skyrmion size and interaction energy on the distance is derived. Two types of behavior are observed depending on the value of iDMI compared to the critical value $D_c$ above which the skyrmion size diverges: For iDMI < $D_c$ the skyrmions acquire a finite size that saturates at maximum value independent of the skyrmion distance that corresponds to the one of the isolated skyrmion. This energy is above that of the homogeneous ferromagnetic state. For iDMI > $D_c$ the skyrmions tend to increase in size to the extent that their neighboring skyrmions permit and finite size skyrmions can be stabilized only in arrays due to their mutual repulsion. These configurations have energy below that of the homogeneous ferromagnetic state.

## 1. Introduction

Democritus proposed that matter, though it appears continuous, it is ultimately composed of indivisible, discrete units called atoms. In an opposite "bottom-up" fashion, skyrmions, as topologically stable structures demonstrate how discrete, particle-like entities can emerge from continuous fields [1]. Magnetic skyrmions as localized, topological spin structures can be viewed as quasiparticles [2]. Magnetic skyrmions can be stabilized at room temperature and zero applied field [3]. Apart from the fundamental scientific interest, arising by the intriguing interplay between the discrete and the continuous in nature, skyrmions rapidly gained intense technological interest as information carriers for future information-processing devices due to their stability, small size and low current densities needed for their transport [4,5,6,7,8]. In particular, the skyrmion racetrack memory (as opposed to the conventional domain wall one) has been proposed as promising ultra-high-density, low-cost and low-power-consumption storage technology [9]. Furthermore, skyrmions can be used in non-conventional computing: A skyrmion-based artificial synapse device for neuromorphic systems has been proposed [10] in which the weight is adjusted based on the interplay between the driving force and the repulsive interactions between skyrmions. In reference [11] a device employing a skyrmion gas to reshuffle a random signal into an uncorrelated copy of itself with possible application to bioinspired stochastic computing is proposed.

Practical applications require the creation, manipulation and detection of isolated skyrmions in magnetic thin-film nanostructures [12]. The skyrmion stability, size and wall width depend sensitively on competing material parameters such as exchange energy, magnetic anisotropy, interfacial Dzyaloshinskii–Moriya interaction (iDMI), and the applied magnetic field [13,14]. There is a range of iDMI values for which isolated skyrmions can exist in thin films. When the iDMI is too weak skyrmions tend to disappear, while for too strong iDMI they increase in size and finally coalesce to a chiral maze-like stripe domain structure. In the in-between range of values metastable isolated skyrmions can be obtained higher than that of the homogeneous ferromagnetic (FM) state [12]. Taking into account the dipolar interactions and under the presence of an external field, bi-stability between two kinds of skyrmions with different sizes is predicted, and skyrmions can be categorized to either stray-field, or DMI stabilized [15]. Through this microscopic-level complexity arising from the coexistence of competing local interactions and non-local stray field interactions, skyrmions emerge at the mesoscopic level as topological quasiparticles which interact with each other as well as with the edges and the possible inhomogeneities of the thin film medium in which they exist. The interaction mechanisms between the skyrmions and between a skyrmion and the sample edge are similar [16,17]. By their own nature, skyrmions permit their description by effective particle models. Usually, an approach proposed by Thiele used [18] where the interactions with the sample edges, defects and between skyrmions is introduced by the gradient of a potential. The interaction of skyrmions at large distances decays exponentially [17,19,20,21,22] within a characteristic magnetic screening length $\lambda$. In skyrmion-based racetrack memories, skyrmions should be located at an appropriate spacing to avoid bit interference and write/read errors. In Ref.[12], a skyrmion bit chain that moves on the racetrack, with the bits at a distance of 57 nm, is reported. The optimization of the spacing between consecutive skyrmionic bits on the racetrack is investigated in Ref[23]. Skyrmionic bits are found to get squeezed together at the end of the racetrack. Such a clogging leads to the reduction of skyrmion size due to their interactions that tend to compress the skyrmions.

In interfacial-DMI systems skyrmions tend to form hexagonal lattices (as the ones of vortices in type II superconductors) due to the repulsive interactions [24, 25]. Each skyrmion in the lattice interacts with its six nearest neighbors. This interaction energy defines the scale for the skyrmion-lattice melting temperature [20]. The skyrmion interaction energies as a function of their distance can be deduced from the properties of such hexagonal lattices having different lattice constants. In this work we study the interactions between skyrmions through the properties of their skyrmion lattices using the mumax3 finite difference micromagnetic simulation program [26,27]. The results for lattices are compared to single isolated skyrmions as a limiting case. The interactions with the film edges are excluded by assuming periodic boundary conditions and the dependence of skymion size and energy on distance is derived. For metastable skyrmions these parameters decay exponentially to the values of the isolated skyrmion. We show that the interactions expand the range of DMI for which isolated skyrmion lattices are observed without dissolving to chiral maze-like domain patterns.

## 2. Micromagnetic simulation details

The following micromagnetic parameters are used: Saturation magnetization $M_S$=580 kA/m, uniaxial anisotropy $K_{mc}$= 800 kJ/m³, exchange stiffness $A_{ex}$=15 pJ/m, and LLG damping α=0.3 [12]. The effective thin film anisotropy is $K = K_{mc} - 0.5\mu_0 M_s^2$ =588 kJ/m³. The results reported here are for iDMI 2.3 to 4 mJ/m²: lower values lead to unstable skyrmions while at higher ones the skyrmions lose their cylindrical symmetry and start to dissolve to domain structures. The micromagnetic parameters give an exchange length $L_{ex} = \sqrt{2A_{ex}/\mu_0 M_s^2}$ as large as 8.4 nm but the spatial discretization must be fine enough for the effects of iDMI: it was set 1 nm and to 0.5 for the DMI values that lead to skyrmions with diameters less than 10nm. The skyrmions considered are Néel-type with positive core i.e. the magnetization vector points upwards (θ=0) at the center and downwards (θ=π) away from the center. The chosen chirality is the one favored by positive iDMI values. We define the size of the skyrmion as the radius of the locus where the angle θ equals to zero. The energy values reported are with respect to that of a negatively magnetized (θ=π) homogeneous ferromagnetic background. To avoid the effects of edge repulsion related to the confinement in finite size film samples periodic boundary conditions have been used. Apart from the spatial wraparound the mumax3 also takes into account several periodic images (copies) in each direction to calculate accurately the effects of the long-range dipolar interactions. We have set this number of copies in each direction to 3 so the dipolar interactions are calculated for a film sample 7×7=49 times larger than the actual cell.

## 3. Isolated skyrmions

The skyrmions can be considered as isolated if their distance is much larger than their size and the screening length, λ that characterizes the exponential decay of their interactions. In the case of a single skyrmion on a film sample the size of the simulation cell must be large enough so that the effect of the interaction with the boundaries (for open boundary conditions) or its own image (for periodic boundary conditions) should be negligible. As a criterion to ensure that a simulation, of a single skyrmion existing on a (L×L) film sample, describes correctly its properties as an isolated entity, the resulting skyrmion size should be small compared to L and independent of L. This criterion should be used for both the interactions with the edges, in the case of open boundary conditions, and the interaction of the skyrmion with its own image, in the case of periodic conditions. For the study of the properties of isolated skyrmions the dimension L was varied from 256 to 2048 nm and periodic boundary conditions were used. As the iDMI value approaches a critical value $D_c$ for which the skyrmion the skyrmion size diverges these conditions never be met within a finite (L×L) film sample and the skyrmion size always adjusts to the available area. Thus, the criterion that skyrmion size should be independent of L is not met and the skyrmion cannot be considered as an isolated entity.

In Figure.1 the size and the energy of an isolated skyrmion are presented as a function of the iDMI value for values less than $D_c$. The energy is defined as the extra per skyrmion energy above that of the homogeneous ferromagnetic state. The data presented here are only for the iDMI for which the size is at most 72nm which is considerably less than the simulation size $L=2048$nm. The results obtained for the skyrmion arrays of the next section, in the limit of large skyrmion separations, are superimposed for comparison.

The results can be interpreted according to the analysis of ref.[14] which uses an analytical model involving the skyrmion size and the wall width of its radial profile as free parameters. Interestingly, and under zero applied external field, following some simple scaling expressions for the exchange, iDMI and anisotropy energy terms one can derive an analytical expression for the skyrmion radius of the form

$$R = \sqrt{\frac{A}{K}} \frac{D}{\sqrt{D_C^2 - D^2}} \qquad (eq.1)$$

which diverges as the iDMI value $D$ approaches the critical value $D_C = 4\sqrt{AK}/\pi$. The micromagnetic simulation data are in fair agreement with this law despite its simplifying underlying assumptions. The fitted values are $D_C = 3.68$ mJ/m$^3$ and $\sqrt{A/K} = 5.9$nm, in fair agreement with the values $D_C = 3.78$ mJ/m$^3$ and $\sqrt{A/K} = 5$ nm derived directly from the used micromagnetic parameters mentioned above. For iDMI>$D_C$ skyrmions cannot exist as isolated entities due to the fact that they expand until they interact with neighboring ones.

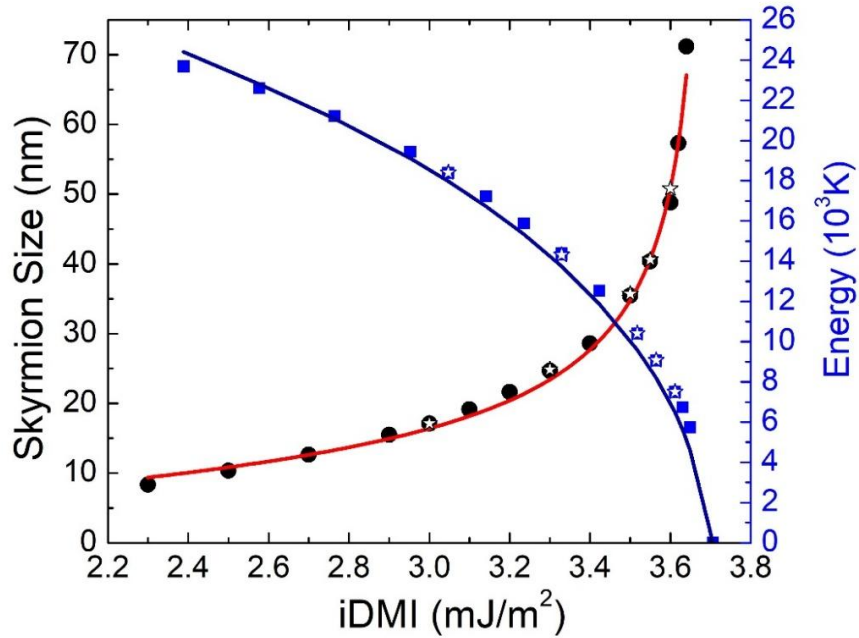

Fig.1 Skyrmion size and energy as a function of the interfacial DMI strength. The energy is given in Klevin calculated as the energy per skyrmion above that of the homogeneous ferromagnetic state. Solid symbols are the results of simulations of single skyrmions presented here, while with stars are denoted the results of simulations of hexagonal lattices in the limit of large skyrmion distances of section 3. The continuous red line is a fit to equation 1. The continuous blue line is the calculated skyrmion energy using the same exactly parameter values derived by the fit of skyrmion size.

## 4. Skyrmion Arrays

In order to cover different limiting cases, with respect to skyrmion separation, and to crosscheck the results, two approaches have been used in parallel:

(i) A state consisting of 25 skyrmions at random initial positions on a film with dimensions ($L\times L$) with $L$ ranging from $L=100$ nm to $L=500$ nm is assumed, and the system is left to relax to its final configuration. When the film dimensions are not sufficient to accommodate all of the 25 skyrmions some of the skyrmions may get annihilated and the final value of the topological charge is $Q_{top}<25$. In these cases, in which the skyrmions become so tightly squeezed that begin to annihilate the results verify that the lattice constant (and skyrmion density thereof) remains the same and does not depend on $L$. For high iDMI values and skyrmion packing fractions the skyrmions start to lose their cylindrical symmetry and become elongated, tending to form maze domains. The results reported here include only the cases for which the skyrmions remain cylindrical and arranged in hexagonal lattices (Fig.2). In the opposite limiting case, in which the film dimensions are large enough so that skyrmion packing fraction is low and interactions become weak the skyrmions do not organize in hexagonal lattices but remain in random positions. For these cases we have used the alternative approach below.

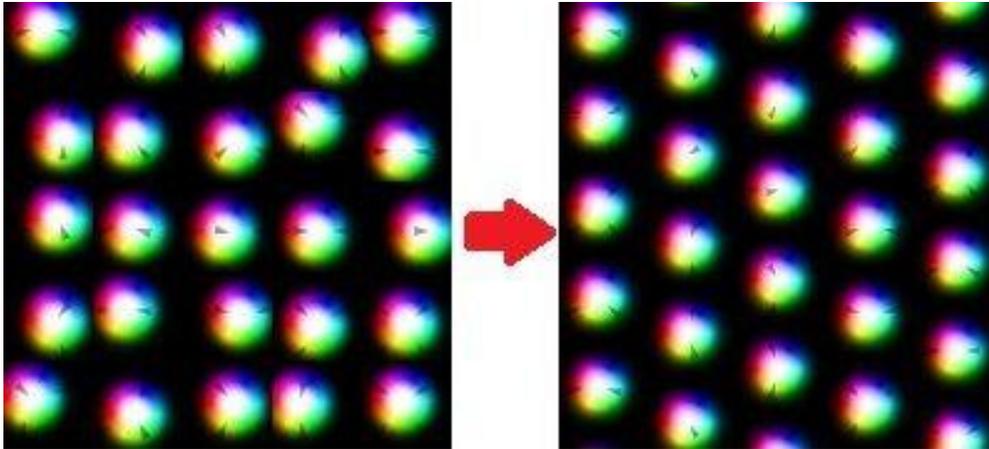

Fig. 2: An array of Néel skyrmions with random initial positions is self-organized to a hexagonal lattice due to their repulsive interactions. The area is $180\times180$nm$^2$. Periodic boundary conditions have been imposed.

(ii) The initial positions of the skyrmions are set on hexagonal 2D-lattices, at nearest neighbor distances $d$, and the lattice is left to relax to their final configuration which is characterized by a specific skyrmion size. We have considered 20 skyrmions arranged in an array of 4 rows of 5 skyrmions on a film with dimensions $5d \times 2\sqrt{3}d$ to make the boundary conditions compatible with the hexagonal symmetry (Fig.3).

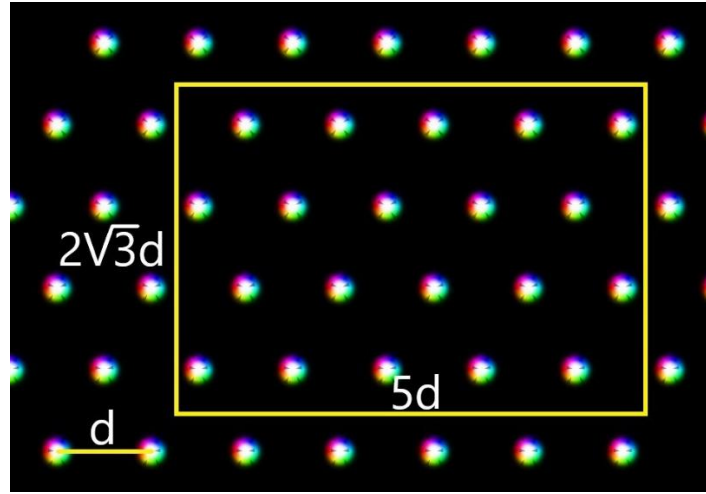

Fig.3 Choice of a rectangular simulation sample with dimensions $5d \times 2\sqrt{3}d$ on which the periodic conditions along the two dimensions can correctly represent a 2D hexagonal array of skyrmions at distances equal to $d$.

In Fig.4 the skyrmion size is plotted as a function of the distance between skyrmions (lattice constant of the 2D-hexagonal cell). For iDMI less than 3.6mJ/m² the size saturates at maximum value independent of the skyrmion distance which corresponds to the isolated skyrmion size derived in the previous section. The data points extrapolate to zero size at a minimum distance ($d_{min}$) close to 10 nm and can be described by sigmoidals of the form

$$s = s_0 \cdot tanh\left(\frac{d - d_{min}}{\lambda_d}\right) \tag{eq.2}$$

where $s_0$ is the size of the isolated skyrmion. Data fitting yields $\lambda_d \approx 1.25 s_0$ and $d_{min} \approx 12$ nm. For stronger iDMI the skyrmions tend to expand and their size is adjusted to the available space. Thus, the parameter $s_0$ cannot be defined. Note that the data presented include only the cases for which the lattice remains hexagonal and the skyrmions retain their cylindrical symmetry, as for values 4 mJ/m² and above the skyrmions deform to complex chiral structures which however conserve the topological charge. The maximum value of size to distance ratio, obtained for iDMI=3.8 and 4 mJ/m², is $s/d = 0.519$. Therefore, the maximum hexagonal lattice packing fraction

$$p = \frac{\pi}{2\sqrt{3}}\left(\frac{s}{d}\right)^2 \tag{eq.3}$$

is close to $p_{max}$=0.244.

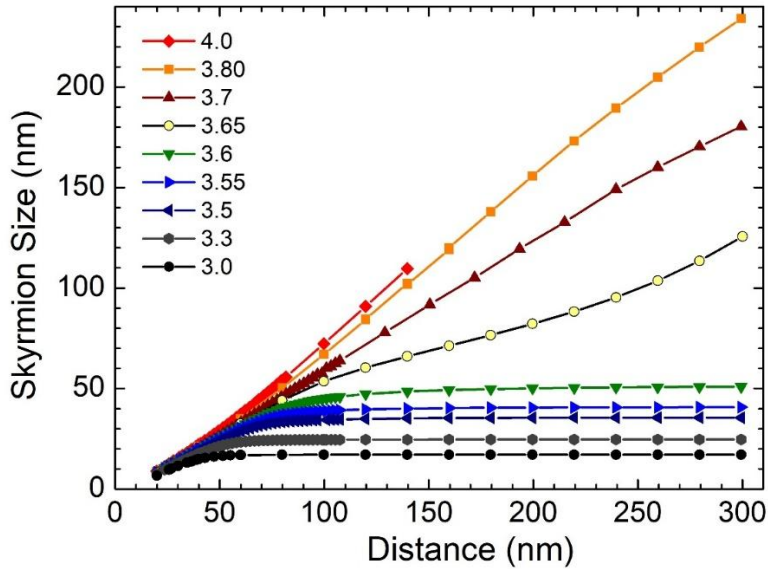

Fig.4 Skyrmion size as a function of the distance between neighboring skyrmions in a hexagonal array for the different values of iDMI indicated (in mJ/m² units).

In Fig.5 the energy per skyrmion is plotted as a function of the distance between skyrmions. For iDMI less than 3.6mJ/m² the energy tends to minimum value independent of the skyrmion distance which corresponds to that of isolated skyrmion size. For these cases the excess energy per skyrmion (above the value of the isolated skyrmion) can be interpreted as the interaction energy $E_{int}$. The data follow an exponential decay law $E(d) = E_{int}e^{-(d-d_{min})/\lambda_E} + E_{iso}$ where $\lambda_E$ is a characteristic screening length, $E_{iso}$ the energy of the isolated skyrmion and $E_{int}$ the interaction energy at the minimum distance $d_{min}$. The $E_{iso}$ data are also superimposed in Fig.1 for comparison.

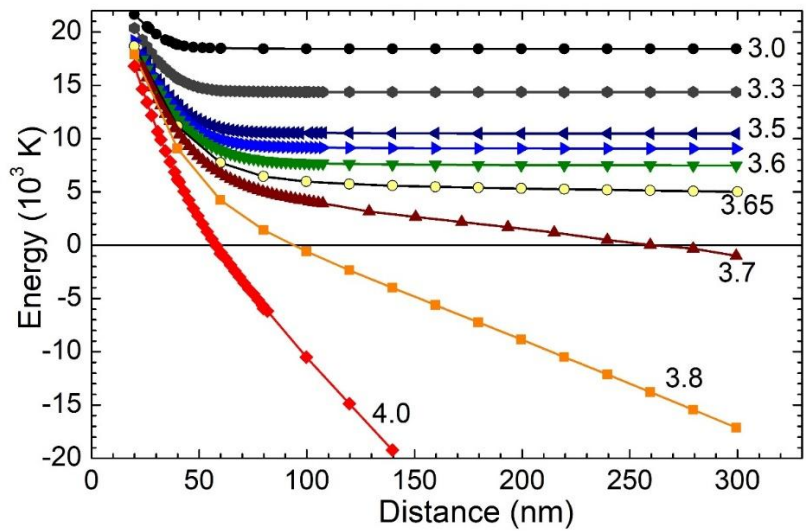

Fig.5 Skyrmion energy as a function of the distance between neighboring skyrmions in a hexagonal array for the different values of iDMI indicated (in mJ/m² units).

The results are summarized in table I below.

TABLE I. Skymrion energies and sizes for different iDMI values

| iDMI (mJ/m3) | $\lambda_E$(nm) | $E_{iso}$ (K) | $E_{int}$ (K) | $E_{iso}(p=0)$ (K) | $E_{int}(p=p_{max})$ (K) |
|---|---|---|---|---|---|
| 3.00 | 10.7 | 18402 | 3216 | 16637 | 1984 |
| 3.30 | 13.0 | 14323 | 6010 | 13354 | 2862 |
| 3.50 | 15.1 | 10420 | 7559 | 10368 | 3870 |
| 3.55 | 16.0 | 9076 | 8652 | 9441 | 4237 |
| 3.60 | 27.1 | 7500 | 9905 | 8397 | 4689 |

For stronger iDMI the skyrmions tend to expand the energy is decreasing with size. Energy minimization leads to an increase of the size up to the extent that other neighboring skyrmions permit. Thus, the energy of the isolated skyrmion size cannot be assessed. The negative energy values show that the skyrmion state is more favorable than that of the homogeneous magnetization, i.e. the skyrmion arrays are thermodynamically stable. For iDMI=3.65 mJ/m² an intermediate behavior is observed and the required simulation times to reach the final equilibrium state increase and approach 15 ns.

## 5. Discussion and conclusions

It is interesting that some insight come be gained, even within the isolated skyrmion model of section 3 (Ref[14]), if the thin film effective anisotropy $K = K_{mc} - (1/2)\mu_0 M_s^2$ is replaced by an expression for a skyrmion array with packing fraction $p$, as $K(p) = K_{mc} - (1/2)\mu_0 M_s^2(1-2p)$. This is justified since under the presence of skyrmions, having opposite magnetization compared to the ferromagnetic background, the magnetization () scales as $M = M_s(1-2p)$ and thus the corresponding demagnetizing field. For instance, at the maximum packing $K = 691.9$ kJ/m³ and $D_C$ increases to 3.9 mJ/m³. Since the $D_C$, which determines the isolated size in arrays, depends itself on the size through the packing fraction, a complex behavior can arise near the critical region of arrays. Of course, dipolar interactions are significant at large separations while at high packing fractions the most important contribution is expected to be the strong repulsive interaction as the skyrmion of exchange origin whirling configurations approach each other. The values of energies determined for $p=0$ and $p=p_{max}=0.244$ are also listed in table I. Indeed, the $E_{iso}$ values are close to those determined for $p=0$ whereas $E_{int}$ values deviate largely from those determined for $p=p_{max}=0.244$. Still, they can give an estimate as to what part of the energy is due to dipolar interactions.

In short, the interactions between skyrmions are deduced through the properties of their hexagonal lattices using the mumax3 finite difference micromagnetic simulation program. Two limiting cases can be distinguished.

(a) For iDMI less than the $D_c$ the skyrmions acquire a finite size that saturates at maximum value independent of the skyrmion distance and corresponds to the one of the isolated skyrmion. This energy is above that of the homogeneous ferromagnetic state. The interaction energy can be deduced as the excess energy per skyrmion (above the value of the isolated skyrmion) and falls off exponentially with the distance between skyrmions within a characteristic is a screening length $\lambda_E$. For distances much larger than $\lambda_E$ the energy becomes independent of distance which means that the inter-skyrmion forces F=$-\partial E/\partial r$ are zero.

(b) For iDMI > $D_c$ the skyrmions tend to increase in size to the extent that other neighboring skyrmions permit. Thus, the energy of the isolated skyrmion of finite size cannot be defined and finite size skyrmions can be stabilized only in arrays due to their mutual repulsion. The energy vs distance curve never flattens off i.e. inter-skyrmion forces persist at all distances due to ever increasing skyrmion size. These configurations have energy below that of the homogeneous ferromagnetic state.